\documentclass[a4paper]{jpconf}
\usepackage{graphicx}
\usepackage{amssymb,amsmath,cite}

\newcommand{\ba}{\begin{eqnarray}}
\newcommand{\ea}{\end{eqnarray}}
\newcommand{\ban}{\begin{eqnarray*}}
\newcommand{\ean}{\end{eqnarray*}}
\newcommand{\bsub}{\begin{subequations}}
\newcommand{\esub}{\end{subequations}}

\def\ket#1{|#1\rangle}
\def\bra#1{\langle#1|}

\begin{document}
\title{Simultaneous occurrence of distinct symmetries\\
in nuclei}

\author{A. Leviatan}

\address{Racah Institute of Physics, The Hebrew University,
Jerusalem 91904, Israel}

\ead{ami@phys.huji.ac.il}

\begin{abstract}
We show that distinct emergent symmetries, such as partial dynamical 
symmetry and quasi dynamical symmetry, can occur simultaneously 
in the same or different eigenstates of the Hamiltonian. 
Implications for nuclear spectroscopy in the rare-earth region 
and for first-order quantum phase transitions between 
spherical and deformed shapes, are considered.
\end{abstract}

\section{Introduction}

Symmetries play an important role in the understanding of complex systems.
However, complete dynamical symmetries (DS) are manifest in only
a few nuclei. Generic Hamiltonians involve interaction terms
with competing incompatible symmetries, which
break the DS. More often one finds that the assumed symmetry is not
obeyed uniformly, {\it i.e.}, is fulfilled by some of the states but not
by others. The need to address such situations, but still preserve important
symmetry remnants, 
has led to the introduction of partial dynamical symmetry 
(PDS)~\cite{Leviatan11}.
For the latter, only part of the eigenspectrum observes a symmetry and 
retains good quantum numbers and/or complete solvability. 
Various types of PDS were proposed
and extensive tests have provided empirical evidence for their relevance
to a broad range of nuclei, 
{\it e.g.}, in the rare-earth~\cite{Lev96,Lev13,Casten14}, 
platinum~\cite{Ramos09,isa15} and actinide~\cite{Couture15} regions, 
in light nuclei~\cite{Escher00} and in semi-magic 
nuclei~\cite{Isacker08}. 
In parallel, the notion of quasi dynamical symmetry (QDS)
was introduced and discussed in the context of nuclear
models~\cite{Bahri00,Rowe04}. 
It expresses the tendency of a Hamiltonian 
to exhibit characteristic properties of the closest DS,
for a certain range of its parameters.
This ``apparent'' (but broken) symmetry,
is due to a coherent mixing of representations 
in selected states, which share a common intrinsic structure.

The concept of PDS reflects the purity of selected states, as opposed
to QDS which reflects a coherent mixing. Up to now, 
these two symmetry concepts have been considered to be unrelated.
In the present contribution, we establish a linkage between them
and show that both can occur simultaneously in the same or in different
eigenstates of the Hamiltonian. Implications 
for the spectroscopy of rare-earth nuclei~\cite{Kremer14}
and for shape-phase transitions~\cite{MacLev11,LevMac12,Macek14} 
are considered.

\section{The interacting boson model: test ground for competing symmetries}

The interacting boson model (IBM)~\cite{ibm}
provides a convenient framework for exploring the role of
symmetries in nuclei. It 
has been widely used to describe quadrupole
collective states in nuclei
in terms of $N$ monopole ($s^\dag$) and quadrupole ($d^\dag$) bosons,
representing valence nucleon pairs.
The model has U(6) as a spectrum generating algebra
and exhibits three DS limits associated with the following 
chains of nested subalgebras,
\bsub
\ba
&&
{\rm U(6)\supset U(5)\supset O(5)\supset O(3)}
\qquad\qquad
\ket{N,\, n_d,\,\tau,\, n_{\Delta},\, L} 
\label{U5}
\\
&&
{\rm U(6)\supset SU(3)\supset O(3)}
\qquad\qquad
\hspace{1.05cm}
\ket{N,\, (\lambda,\mu),\,K,\, L} 
\label{SU3}
\\
&&
{\rm U(6)\supset O(6) \supset O(5)\supset O(3)}
\qquad\qquad
\ket{N,\, \sigma,\,\tau,\, n_{\Delta}, L} ~.
\label{O6}
\ea
\esub
The indicated basis states are classified by quantum numbers which are the
labels of irreducible representation (irreps) of the algebras in each
chain. These solvable limits 
correspond to known benchmarks of
the geometric description of nuclei, 
involving vibrational [U(5)], $\gamma$-soft [O(6)], 
and rotational [SU(3)] types of dynamics. 
This identification is consistent with the
geometric visualization of the model. The latter is
obtained by an energy surface, $E(\beta,\gamma)$, 
defined by the expectation value of the Hamiltonian in the coherent
(intrinsic) state~\cite{gino80},
\ba
\vert\beta,\gamma ; N \rangle &=&
(N!)^{-1/2}(b^{\dagger}_{c})^N\,\vert 0\,\rangle ~,
\label{condgen}
\ea
where, $b^{\dagger}_{c} \propto \beta\cos\gamma
d^{\dagger}_{0} + \beta\sin{\gamma}
( d^{\dagger}_{2} + d^{\dagger}_{-2})/\sqrt{2} + s^{\dagger}$.
Here $(\beta,\gamma)$ are
quadrupole shape parameters whose values, $(\beta_{\rm eq},\gamma_{\rm eq})$,
at the global minimum of $E(\beta,\gamma)$ define the equilibrium
shape for a given Hamiltonian. 
For one- and two-body interactions, the  energy surface is of the form
\ba
E(\beta,\gamma) \propto
(1+\beta^2)^{-2}\beta^2
\left[a-b\beta\cos{3\gamma}+c\beta^2\right] ~.
\label{enesurf}
\ea
The shape can be spherical $(\beta =0)$ or
deformed $(\beta >0)$ with $\gamma =0$ (prolate),
$\gamma =\pi/3$ (oblate), 
or $\gamma$-independent. 
The equilibrium deformations associated with the 
DS limits are
$\beta_{\rm eq}=0$ for U(5), $(\beta_{\rm eq} =\sqrt{2},\gamma_{\rm eq}=0)$
for SU(3) and $(\beta_{\rm eq}=1,\gamma_{\rm eq}\,{\rm arbitrary})$ for O(6).
\begin{figure}[t]
\includegraphics[width=17.7pc]{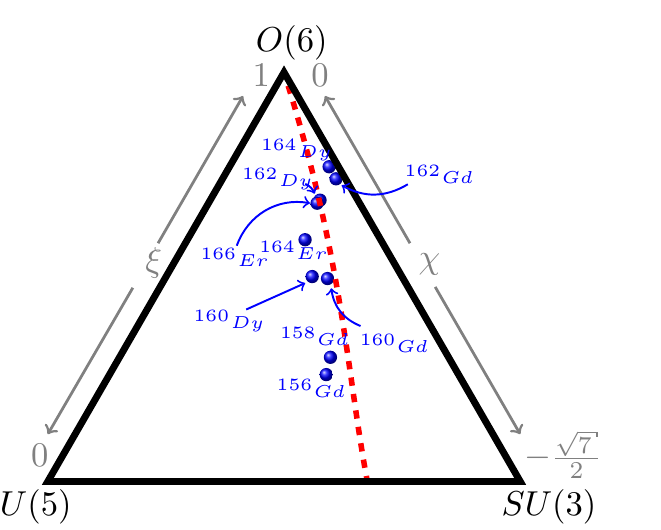}\hspace{2pc}%
\begin{minipage}[b]{20pc}
\caption{
\small
The ECQF symmetry triangle with the position of
the rare-earth nuclei of Table~\ref{nuclei} indicated by bullets. 
The calculated (for $N\!=\!60$) red dashed line, 
correspond to a region of an approximate ground-state 
O(6) symmetry, exemplifying an O(6)-PDS, as discussed in Section~3. 
From~\cite{Kremer15}.}
\label{fig:triangle}
\end{minipage}
\end{figure}

The extended consistent-Q formalism (ECQF)~\cite{Warner83} of the IBM, 
uses the same quadrupole operator, 
$\hat Q^\chi=d^\dag s+s^\dag\tilde d+\chi\,(d^\dag\tilde d)^{(2)}$,
in the E2 transition operator
and in the Hamiltonian,
\ba
\hat H_{\rm ECQF}=
\omega\left[(1-\xi)\,\hat n_d-\frac{\xi}{4N}\,
\hat Q^\chi\cdot\hat Q^\chi\right] ~.
\label{eq:Hamiltonian}
\ea
Here $\hat n_d$ is the d-boson number operator, 
$\tilde{d}_{m}=(-)^{m}d_{-m}$
and the dot implies a scalar product.
$\xi$~and $\chi$ are the sole structural parameters of the model
since $\omega$ is a scaling factor.
The parameter ranges $0\leq\xi\leq1$ and
$-\frac{\sqrt{7}}{2}\leq\chi\leq0$
interpolate between the U(5), O(6) and SU(3) DS limits,
which are reached for
$(\xi,\chi)=(0,\chi)$, $(1,0)$, and $(1,-\frac{\sqrt{7}}{2})$,
respectively.
It is customary to represent the parameter space by a symmetry
triangle~\cite{Casten83}, shown in Fig.~1,
whose vertices correspond to these limits.
The ECQF has been used extensively~for 
describing nuclei~\cite{McCutchan04} 
and it was found that the vast majority of nuclei
are best described by ECQF parameters in the interior of the triangle,
away from any DS limit. 
In what follows we examine
the O(6) symmetry properties of ground-band states in such nuclei,
in the rare-earth region, using $\hat H_{\rm ECQF}$~(\ref{eq:Hamiltonian}).

\section{
Simultaneous occurrence of
O(6)-PDS and SU(3)-QDS in rotational nuclei}

Given an eigenstate $\ket{L}$, with
angular momentum $L$, of the ECQF Hamiltonian~(\ref{eq:Hamiltonian}),
its expansion in the O(6) DS basis reads
$\ket{L} = \sum_i{C_i\,\ket{N,\sigma_i,\tau_i,L}}$. 
The degree of O(6) symmetry of the state $\ket{L}$ 
is inferred from the fluctuations in $\sigma$, 
calculated as~\cite{Kremer14},
\ba
\Delta\sigma_L=
\sqrt{\sum_i{C_i^2\,\sigma_i^2}
-\left(\sum_i{C_i^2\,\sigma_i}\right)^2} ~.
\label{eq:fluc2}
\ea
If $\ket{L}$ carries an exact O(6) quantum number,
$\sigma$ fluctuations are zero, $\Delta\sigma_L=0$.
If $\ket{L}$ contains basis states with different O(6) quantum numbers,
then $\Delta\sigma_L>0$, indicating that the O(6) symmetry is broken.
Note that $\Delta\sigma_L$ also vanishes
for a state with a mixture of components with the same $\sigma$
but different O(5) quantum numbers $\tau$,
corresponding to a state $\ket{L}$ with good O(6) but mixed O(5) character.
$\Delta\sigma_L$ has the same physical content as wave-function entropy.

\begin{figure}[t]
\begin{minipage}{17.5pc}
\includegraphics[width=17.5pc]{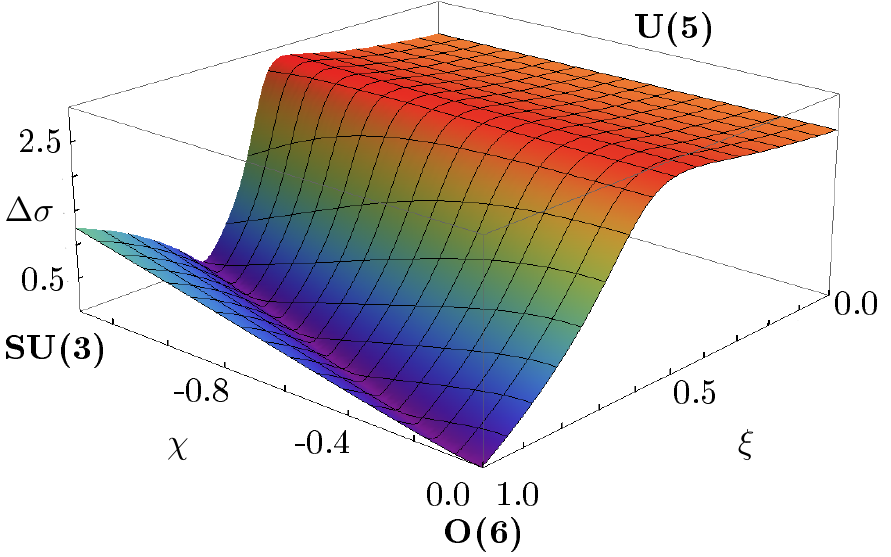}
\caption{
\small
Ground-state ($L\!=\!0^{+}_1$)~fluctuations
$\Delta\sigma_{0}$~(\ref{eq:fluc2})
for $\hat{H}_{\rm ECQF}$~(\ref{eq:Hamiltonian})
with $N\!=\!14$.
The fluctuations vanish at the O(6) DS limit,
saturate towards the U(5) DS limit,
and are of the order $10^{-2}$ in the valley.
Adapted from~\cite{Kremer14}.}
\label{fig:3d}
\end{minipage}\hspace{2pc}%
\begin{minipage}{17pc}
\vspace{0.35cm}
\includegraphics[width=17pc]{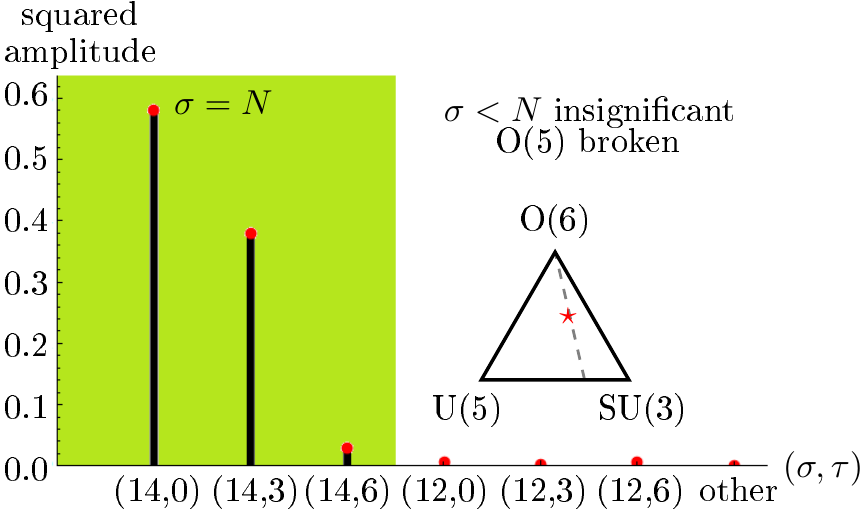}
\caption{
\small
Squared amplitudes $C_i^2$ in the expansion in the O(6)
basis~(\ref{O6}), of the $L\!=\!0^{+}_1$ ground state of
the ECQF Hamiltonian~(\ref{eq:Hamiltonian})
with parameters $\xi\!=\!0.84,\,\chi\!=\!-0.53,\,N\!=\!14$,
appropriate for $^{160}$Gd. Adapted from~\cite{Kremer14}.}
\label{fig:wavefunction}
\end{minipage}
\end{figure}

The fluctuations $\Delta\sigma_L$ can now be examined
for the entire parameter space of the ECQF
Hamiltonian~(\ref{eq:Hamiltonian}).
Results of this calculation for the ground state,
$\ket{L=0^{+}_1}$, with $N=14$,
are shown in Fig.~\ref{fig:3d}.
At the O(6) DS limit ($\xi=1$, $\chi=0$),
$\Delta\sigma_{0}\equiv\Delta\sigma_{L=0_1}$ vanishes per construction
whereas it is greater than zero for all other parameter pairs.
Towards the U(5) DS limit ($\xi=0$),
the fluctuations reach a saturation value of
$\Delta\sigma_{0}\approx 2.47$.
At the SU(3) DS limit ($\xi=1$, $\chi=-\frac{\sqrt{7}}{2}$)
the fluctuations are $\Delta\sigma_{0}\approx 1.25$.
Surprisingly, one recognizes a valley
of almost vanishing $\Delta\sigma_{0}$ values,
two orders of magnitude lower than at saturation.
This region (depicted by a red dashed arc in the triangle of
Fig.~\ref{fig:triangle}),
represents a parameter range of the IBM,
outside the O(6) DS limit,
where the ground-state wave function
exhibits an exceptionally high degree of purity
with respect to the O(6) quantum number $\sigma$.
\begin{table}[t]
\caption{
\small
Calculated $\sigma$ fluctuations
$\Delta\sigma_{L}$, Eq.~(\ref{eq:fluc2}),
for rare earth nuclei in the vicinity of the identified region of
approximate ground-state-O(6) symmetry~\cite{Kremer14}.
Also shown are the fraction
$f^{(L)}_{\rm \sigma=N}$ of O(6) basis states with $\sigma = N$ contained
in the $L\!=\!0,2,4$ states, members of the ground band.
The structure parameters $\xi$ and $\chi$ of the 
ECQF Hamiltonian (\ref{eq:Hamiltonian}) employed, are taken
from \cite{McCutchan04}.\label{nuclei}}
\begin{center}
\lineup
\begin{tabular}{lccccccccc}
\br
Nucleus & $N$ & $\xi$ & $\chi$
& $\Delta\sigma_{0}$ & $f^{(0)}_{\rm \sigma=N}\;\;$
& $\Delta\sigma_{2}$ & $f^{(2)}_{\rm \sigma=N}\;\;$
& $\Delta\sigma_{4}$ & $f^{(4)}_{\rm \sigma=N}$\cr
\mr
$^{156}$Gd & 12 & 0.72 & -0.86 & 0.46 & 95.3\% &
0.43 & 95.8\% & 0.38 & 96.6\% \\
$^{158}$Gd & 13 & 0.75 & -0.80 & 0.35 & 97.2\% &
0.33 & 97.5\% & 0.30 & 97.9\% \\
$^{160}$Gd & 14 & 0.84 & -0.53 & 0.19 & 99.1\% &
0.19 & 99.2\% & 0.17 & 99.3\% \\
$^{162}$Gd & 15 & 0.98 & -0.30 & 0.17 & 99.3\% &
0.17 & 99.3\% & 0.16 & 99.3\% \\
$^{160}$Dy & 14 & 0.81 & -0.49 & 0.44 & 96.2\% &
0.39 & 96.4\% & 0.36 & 96.8\% \\
$^{162}$Dy & 15 & 0.92 & -0.31 & 0.07 & 99.9\% &
0.07 & 99.9\% & 0.06 & 99.9\% \\
$^{164}$Dy & 16 & 0.98 & -0.26 & 0.13 & 99.6\% &
0.13 & 99.6\% & 0.13 & 99.6\% \\
$^{164}$Er & 14 & 0.84 & -0.37 & 0.39 & 96.5\% &
0.37 & 96.7\% & 0.35 & 97.1\% \\
$^{166}$Er & 15 & 0.91 & -0.31 & 0.12 & 99.7\% &
0.11 & 99.7\% & 0.10 & 99.7\% \\
\br
\end{tabular}
\vspace{-12pt}
\end{center}
\end{table}

The ground-state wave functions in the valley of low $\Delta\sigma_{0}$,
can be expanded in the O(6)-DS basis (\ref{O6}).
At the O(6) DS limit only one O(6) basis state, with $\sigma=N$ and
$\tau=0$ contributes, while outside this limit
the wave function consists of multiple O(6) basis states.
Investigation of the wave function for parameter combinations inside
the valley reveals an overwhelming dominance of the O(6) basis states
with $\sigma=N$.
This is seen in Fig.~\ref{fig:wavefunction} for the ground-state
wave function of $\hat{H}_{\rm ECQF}$~(\ref{eq:Hamiltonian}),
with parameter values that apply to the nucleus $^{160}$Gd. 
The $\sigma=N$ states comprise
more than 99\% of the ground-state wave function
at the bottom of the valley
and their dominance causes $\Delta\sigma_{0}$ to be small.
At the same time, the O(5) symmetry is broken,
as basis states with different quantum number $\tau$
contribute significantly to the wave function.
Consequently, the valley can be identified
as an entire region in the symmetry triangle
with an approximate O(6)-PDS, 
which means that some of the eigenstates exhibit some of the symmetries 
in the chain~(\ref{O6}). 
Outside this valley the ground state is a mixture of several $\sigma$ values
and $\Delta\sigma_{0}$ increases.

Detailed ECQF fits for energies and electromagnetic
transitions of rare-earth nuclei~\cite{McCutchan04}, 
allow one to relate the structure of collective nuclei
to the parameter space of the ECQF Hamiltonian~(\ref{eq:Hamiltonian}).
From the extracted ($\xi,\chi$) parameters one can calculate
the fluctuations $\Delta\sigma_L$
and the fractions $f_{\sigma=N}$ of squared $\sigma=N$ amplitude.
Nuclei with $\Delta\sigma_{0} < 0.5$ and $f_{\sigma=N}>95\%$ in the
ground-state ($L=0^{+}_1$) are listed in Table~\ref{nuclei}.
These quantities are also calculated for yrast states
with $L>0$ and exhibit similar values in each nucleus.
It is evident that a large set of rotational rare earth nuclei 
are located in the valley of small $\sigma$ fluctuations.
They can be identified as candidate nuclei
with an approximate O(6)-PDS not only for the ground state,
but also for the members of the band built on top of it. 

\begin{figure}[t]
\centering
\includegraphics[width=16cm]{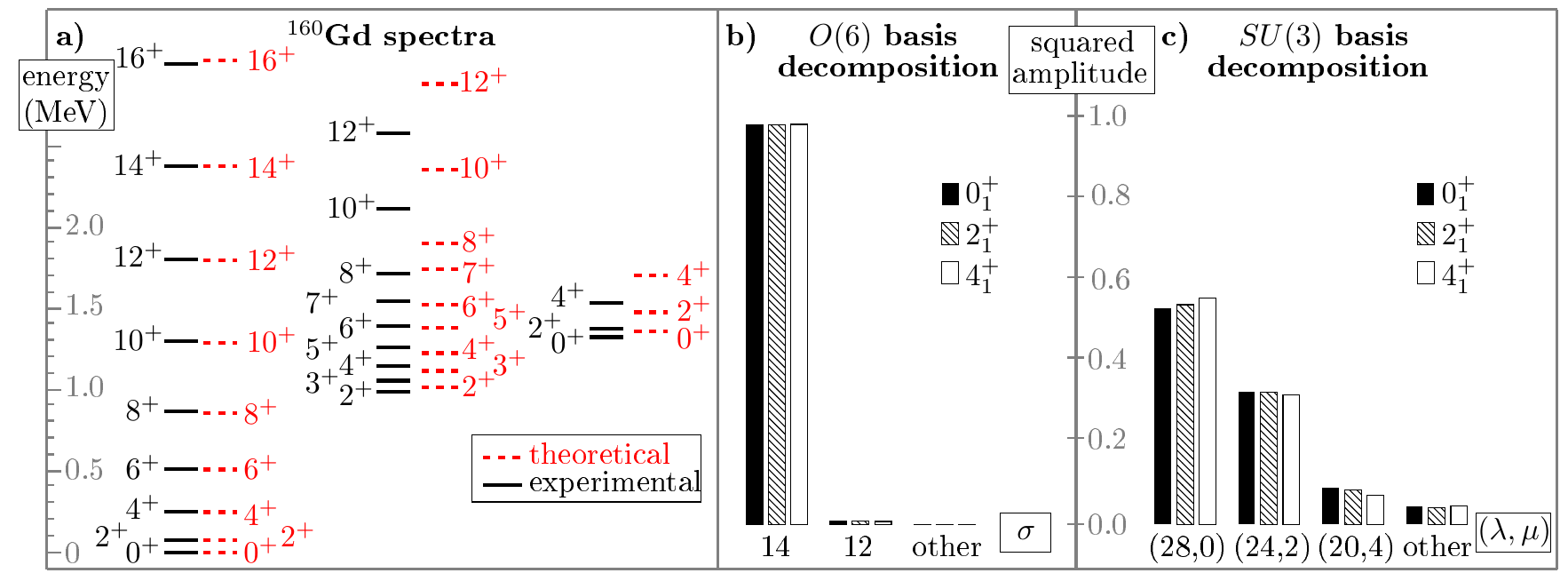}
\caption{
\small
a)~The experimental spectrum of $^{160}$Gd
compared with the IBM calculation
using the ECQF Hamiltonian~(\ref{eq:Hamiltonian})
with parameters $\xi=0.84$ and $\chi=-0.53$
taken from Ref.~\cite{McCutchan04}.
b)~The O(6) decomposition in $\sigma$ components of yrast states
with $L=0,2,4$.
c)~The SU(3) decomposition in $(\lambda,\mu)$ components of
the same yrast states. Adapted from~\cite{Kremer14}.}
\label{fig:160Gd}
\end{figure}

The experimental spectrum of a representative nucleus from 
Table~\ref{nuclei}, $^{160}$Gd, along with its ECQF 
description~(\ref{eq:Hamiltonian}),
is shown in Fig.~\ref{fig:160Gd}(a). 
Figures~\ref{fig:160Gd}(b) and~\ref{fig:160Gd}(c) 
show the decomposition into O(6) and SU(3) basis states, respectively,
for yrast states with $L=0,2,4$.
It is evident that the SU(3) symmetry is broken,
as significant contributions of basis states
with different SU(3) quantum numbers $(\lambda,\mu)$ occur.
It is also clear from Fig.~\ref{fig:160Gd}(c)
that this mixing occurs in a coherent manner
with similar patterns for the different members of the ground-state band.
Such coherent mixing is the hallmark of SU(3) QDS~\cite{Rowe04}. 
On the other hand, as seen in Fig.~\ref{fig:160Gd},
the same yrast states with $L=0,2,4$
are almost entirely composed out of O(6) basis states with 
$\sigma\!=\!N=\!14$
which implies small fluctuations $\Delta\sigma_L$ (\ref{eq:fluc2})
and the preservation of O(6) symmetry (but with broken O(5) symmetry) 
in the ground-state band. 
Thus an empirically-manifested link is established between SU(3) QDS and
O(6) PDS.

The simultaneous occurrence of SU(3)-QDS and O(6)-PDS 
for members of the ground band, signals the existence of a single 
intrinsic state with good O(6) character. Such an intrinsic state is the 
condensate of Eq.~(\ref{condgen}) with $(\beta=1,\gamma=0)$, that has 
$\sigma=N$. The states projected from it 
keep exact O(6) symmetry ($\sigma=N$)
but break the O(5) symmetry (mixed $\tau$)
and have a high overlap with the yrast eigenstates 
of $\hat H_{\rm ECQF}$ 
(more than 99\% for the ground states $L=0^{+}_1$). 
This suggests that the indicated intrinsic state 
provides a good approximation, in a variational sense,
to the ground band of $\hat H_{\rm ECQF}$ along the valley of 
low-$\Delta\sigma_0$. 
The two extremum equations for $E(\beta,\gamma)$, Eq.~(\ref{enesurf}),
$\partial E/\partial\beta=\partial E/\partial\gamma=0$,
have $\beta=1$ and $\gamma=0$ as a solution, provided $b=2c$.
For large $N$, the energy surface coefficients of 
$\hat H_{\rm ECQF}$ are $b=-\omega\xi\sqrt{\frac{2}{7}}\chi/N$ and
$c=\omega\left[1-\xi-\xi\chi^2/14\right]/N$.
Thus, in the valley of low $\Delta\sigma_0$
the desired condition, $b=2c$, fixes $\xi$ to be
$\xi=1/[1-\chi/\sqrt{14}+\chi^2/14]$. 
This relation predicts the location of the region of approximate
ground-state O(6) symmetry for large $N$ very precisely~\cite{Kremer14}. 
These results demonstrate that
coherent mixing of one symmetry (QDS)
can result in the purity of a quantum number
associated with partial conservation of a different, incompatible symmetry
(PDS).

\section{
Simultaneous occurrence of
U(5)-PDS and SU(3)-QDS in shape-coexistence}

\begin{figure}[t]
\begin{minipage}{12.3pc}
\vspace{1.4cm}
\includegraphics[width=12.3pc]{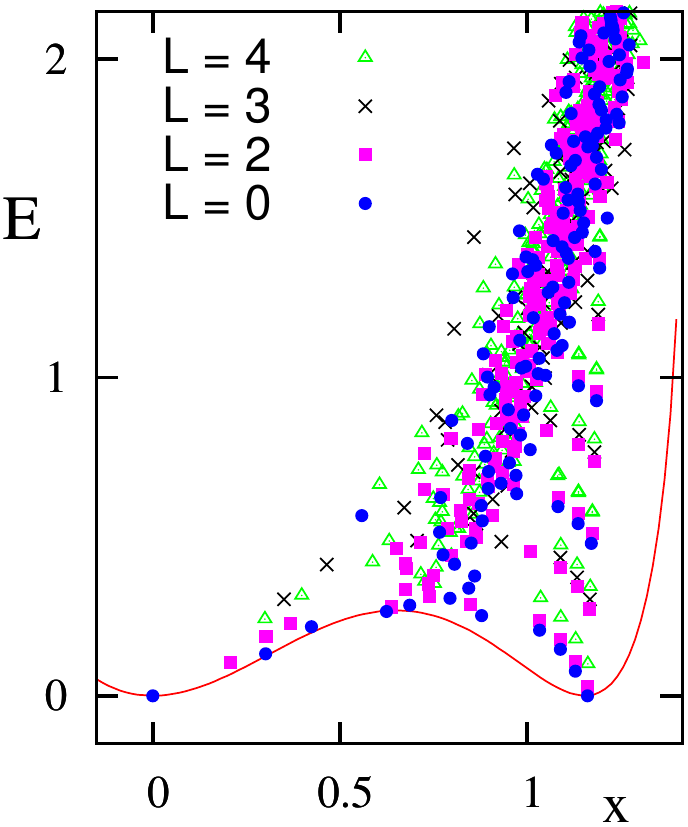}
\caption{
\small
Peres lattices $\{x_i,E_i\}$ for $L\!=\!0,2,3,4$ and $N\!=\!50$
eigenstates of $\hat{H}_{\rm cri}$, Eq.~(\ref{Hcri}),
overlayed on the classical potential. 
Here $x_i \equiv \sqrt{2 \bra{L_i}\hat{n}_d\ket{L_i}/N}$. 
Adapted from~\cite{Macek14}.}
\label{fig:Peres}
\end{minipage}\hspace{2pc}%
\hspace{-0.65cm}
\begin{minipage}{25pc}
\includegraphics[width=12.5pc]{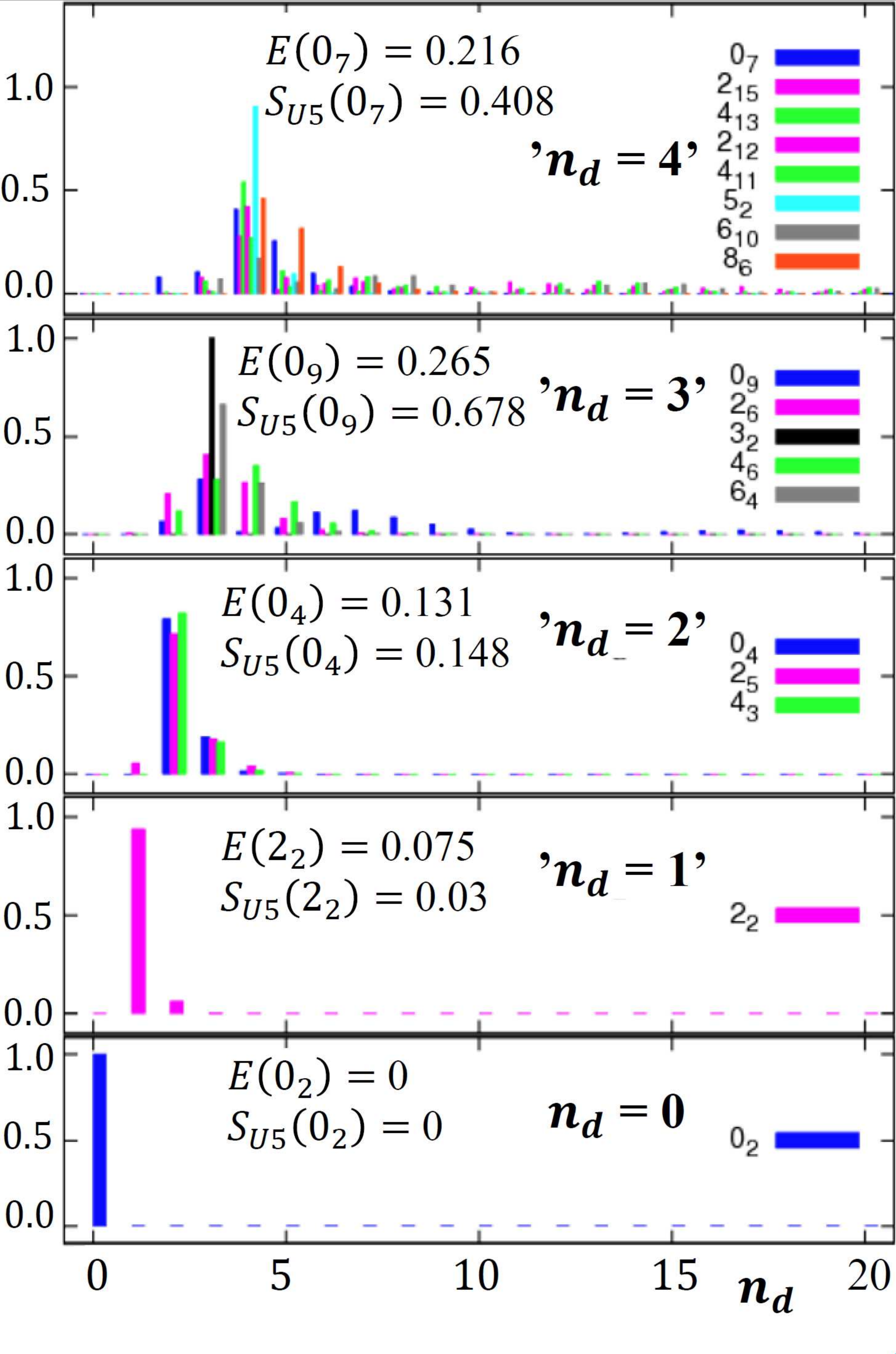}
\includegraphics[width=12.5pc]{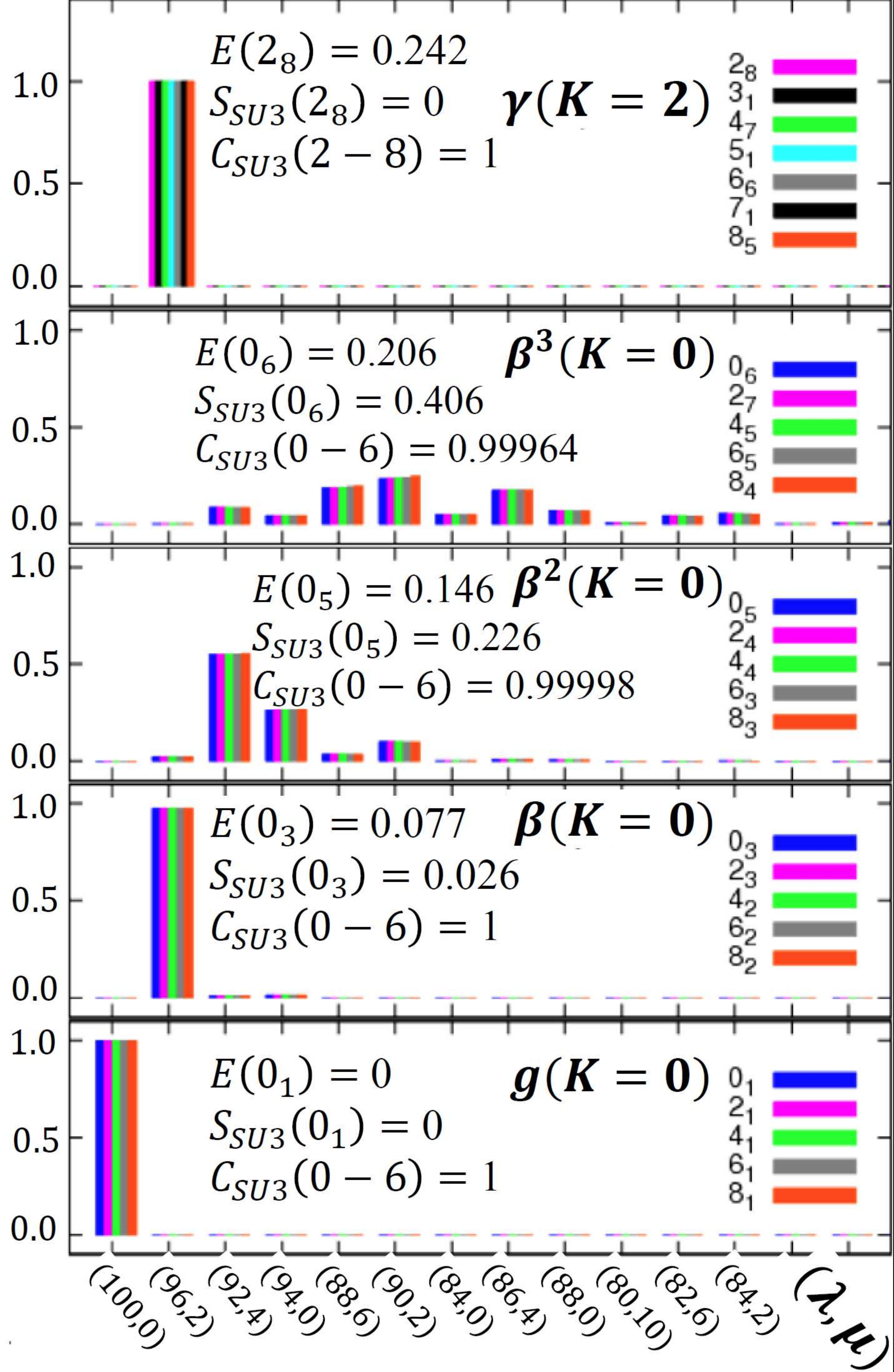}
\caption{
\small
Left column: 
U(5) $n_d$-probability distribution for spherical 
type of eigenstates of $\hat{H}_{\rm cri}$, Eq.~(\ref{Hcri}),
arranged in U(5)-like $n_d$-multiplets. 
Right column: 
SU(3) $(\lambda,\mu)$-probability distribution 
for deformed type of eigenstates arranged in rotational $K$-bands. 
Shannon entropies $S_{\rm U5}(L_i)\!\approx\! 0$ and 
$S_{\rm SU3}(L_i)\!\approx\! 0$
signal U(5)-PDS and SU(3)-PDS, respectively.
SU(3) Pearson correlator $C_{\rm SU3}(0{\rm -}6)\!\approx\!1$,
signals SU(3)-QDS. Adapted from~\cite{Macek14}.}
\label{fig:u5su3}
\end{minipage}
\end{figure}

PDS (partial purity) and QDS (coherent mixing) of distinct symmetries 
can occur simultaneously also in different sets of eigenstates of 
a given Hamiltonian. 
Such a situation is encountered, for example, in a first-order quantum 
phase transition (QPT) involving shape coexistence. 
Focusing on the intrinsic dynamics~\cite{lev87}
at the critical-point such QPT 
between spherical [U(5)] and deformed [SU(3)] shapes, the relevant 
IBM Hamiltonian annihilates the corresponding condensates 
(\ref{condgen}) with $\beta=0$ and $(\beta=\sqrt{2},\gamma=0)$, and 
can be transcribed in the
form~\cite{Lev06,Lev07}
\begin{equation}
\hat{H}_{\rm cri}=
h_{2}\,P^{\dagger}_{2}\cdot \tilde{P}_{2}\;\;\;,\;\;\;
P^{\dagger}_{2m} = 2d^{\dagger}_{m}s^{\dagger} +
\sqrt{7}\, (d^{\dagger}\,d^{\dagger})^{(2)}_{m} \;\;\; , \;\;\;
\tilde P_{2m} = (-)^{m}P_{2,-m} ~.
\label{Hcri}
\end{equation}
$\hat{H}_{\rm cri}$ mixes terms from different DS chains of the IBM,
hence is non-integrable. The corresponding classical Hamiltonian,
obtained by Glauber coherent states, has a Landau potential with two
degenerate spherical and prolate-deformed minima, as shown 
in Fig.~\ref{fig:Peres}. A detailed classical 
analysis~\cite{MacLev11,LevMac12,Macek14} reveals 
a robustly regular dynamics in the region of the deformed minimum 
and a change with energy from regular to chaotic dynamics 
in the region of the spherical minimum.
The mixed but well separated dynamics persists even 
at energies far exceeding the barrier height. 
A quantum analysis is based on Peres lattices~\cite{Peres84}, 
which are constructed by plotting the expectation 
values $x_i \equiv \sqrt{2 \bra{i}\hat{n}_d\ket{i}/N}$
of the $d$-boson number operator,
versus the energy $E_i = \bra{i}\hat{H}\ket{i}$
of the Hamiltonian eigenstates $\ket{i}$.
The lattices $\{x_i,E_i\}$ corresponding to
regular dynamics display an ordered pattern, while chaotic
dynamics leads to disordered meshes of points~\cite{Peres84}.
The quantity $x_i$ is related to the coordinate $x$ in the classical
potential, hence the indicated lattices can distinguish regular from
irregular states and associate them with a given region in phase space.

The Peres lattices for eigenstates of $\hat{H}_\mathrm{cri}$~(\ref{Hcri}), 
with $L\!=\!0,2,3,4$, are shown in Fig.~\ref{fig:Peres}, 
overlayed on the classical potential. They disclose 
regular sequences of states localized within and above the deformed well. 
They are comprised of rotational states with $L=0,2,4,\ldots$
forming regular $K\!=\!0$ bands and sequences $L=2,3,4,\ldots$
forming $K=2$ bands.
Additional $K$-bands (not shown in Fig.~\ref{fig:Peres}),
corresponding to multiple $\beta$ and $\gamma$ vibrations
about the deformed shape, can also be identified.
Such ordered band-structures persist to energies above the barrier and
are not present in the disordered (chaotic) portions
of the Peres lattice.
At low-energy, in the vicinity of the spherical well, one can also detect
multiplets of states with $L=0$, $L=2$ and $L=0,2,4$, typical of
quadrupole excitations with $n_d=0,1,2,$ of a spherical shape.

The nature of the surviving regular sequences of selected states, 
is revealed in a symmetry analysis of their wave functions.
The left column of Fig.~\ref{fig:u5su3} shows the U(5) 
$n_d$-probabilities, $P_{n_d}^{(L_i)}$, for eigenstates of 
$\hat{H}_{\rm cri}$ (\ref{Hcri}), selected on the 
basis of having the largest components with $n_d=0,1,2,3,4$,
within the given $L$ spectra.
The states are arranged into panels labeled
by `$n_d$' to conform with the structure of the $n_d$-multiplets of the
U(5) DS limit. 
In particular, the zero-energy $L\!=\!0^{+}_2$ state
is seen to be a pure $n_d\!=\!0$ state,
with vanishing U(5) Shannon entropy, $S_{\rm U5}\!=\!0$.
It is a solvable eigenstate of $\hat{H}_{\rm cri}$,
exemplifying U(5)-PDS~\cite{Lev07}.
The state $2^{+}_2$ has a pronounced $n_d\!=\!1$
component~(96\%) and the states ($L=0^{+}_4,\,2^{+}_5,\,4^{+}_3$)
in the third panel, have a pronounced $n_d\!=\!2$ component and
a low value of $S_{\rm U5}< 0.15$.
All the above states with $`n_d\leq 2$' have a dominant single $n_d$
component, and hence qualify as `spherical' type of states.
They are the left-most states 
in the Peres lattices of Fig.~\ref{fig:Peres}.
In contrast, the states in the panels `$n_d=3$' and `$n_d=4$' of 
Fig.~\ref{fig:u5su3}, are significantly fragmented. A notable
 exception is the $L=3^{+}_2$ state, which is a solvable U(5)-PDS 
eigenstate with $n_d=3$~\cite{Lev07}.
The existence in the spectrum of specific spherical-type of states with
either $P_{n_d}^{(L)}\!=\!1$ $[S_{\rm U5}(L)\!=\!0]$
or $P_{n_d}^{(L)}\approx 1$ $[S_{\rm U5}(L)\approx 0]$, exemplifies
the presence of an exact or approximate U(5) PDS at the critical-point.

The states considered in the right column of Fig.~\ref{fig:u5su3}
have a different character. They
belong to the five lowest regular sequences
seen in the Peres lattices of Fig.~\ref{fig:Peres}, 
in the region $x\geq 1$. 
They have a broad $n_d$-distribution, 
hence are qualified as `deformed'-type of states,
forming rotational bands:
$g(K\!=\!0),\,\beta(K\!=\!0),\,\beta^2(K\!=\!0),
\,\beta^3(K\!=\!0)$ and $\gamma(K\!=\!2)$. 
The ground $g(K\!=\!0)$ and $\gamma(K\!=\!2)$
bands are pure [SU(3) $(\lambda,\mu)$-distribution, 
$P_{(\lambda,\mu)}^{(L)}=1$, and SU(3) Shannon entropy, $S_{\rm SU3}(L)=0$] 
with $(\lambda,\mu) = (2N,0)$ and $(2N-4,2)$ SU(3) character,
respectively. These are solvable bands of $\hat{H}_{\rm cri}$ and
exemplify SU(3)-PDS~\cite{Lev07}.
The non-solvable $K$-bands are mixed with respect to SU(3)
in a coherent, $L$-independent manner, hence exemplify SU(3)-QDS.
As expected, the Pearson correlator~\cite{Macek10} 
is $C_{\rm SU3}(0_i{\rm -}6)\approx 1$,  
for these regular $K$-bands. 
The above results demonstrate that PDS and QDS can characterize the 
remaining regularity in a system, 
amidst a complicated (at time chaotic) environment of other states.

\section{Concluding remarks}

Both PDS and QDS do not arise from invariance properties of the 
Hamiltonian, hence can occur simultaneously in atomic nuclei. 
As shown, the existence of a region of almost exact ground-state-band
O(6) symmetry outside the O(6) DS limit of the IBM, 
can be understood in terms of an approximate O(6) PDS.
The same wave functions display coherent ($L$-independent) mixing of 
SU(3) irreps and hence comply with the conditions of an 
SU(3) QDS. Many rare-earth nuclei do exhibit this linkage.
Both types of emergent symmetries 
can characterize persisting regular patterns in nuclei, 
{\it e.g.}, U(5)-like and SU(3)-like multiplets, 
amidst a complicated environment of other states, 
a situation encountered in a quantum shape-phase transition. 

The work reported in Section~3, 
was done in collaboration with C. Kremer, 
J. Beller, N. Pietralla, R. Trippel (Darmstadt), G. Rainovski (Sofia), 
P. Van Isacker (GANIL) and in Section~4, with M. Macek (Yale), 
and is supported by the Israel Science Foundation.

\end{document}